\documentclass[fleqn,10pt]{wlscirep}
\usepackage{subfigure}
\usepackage[utf8]{inputenc}
\usepackage[T1]{fontenc}
\title{A complete continuous-variable quantum computation architecture based on the 2D spatiotemporal cluster state}

\author[1,2]{Peilin Du}
\author[1,2,3*]{Jing Zhang}
\author[1,2,3]{Tiancai Zhang}
\author[1,2,3*]{Rongguo Yang}
\author[1,2,3]{Jiangrui Gao}
\affil[1]{State Key Laboratory of Quantum Optics Technologies and Devices, Shanxi University, Taiyuan, 030006, China}
\affil[2]{College of Physics and Electronic Engineering, Shanxi University,Taiyuan, 030006, China}
\affil[3]{Collaborative Innovation Center of Extreme Optics, Shanxi University, Taiyuan, 030006, China}

\affil[*]{zjj@sxu.edu.cn; yrg@sxu.edu.cn}

\begin{abstract}
Continuous-variable measurement-based quantum computation, which requires deterministically generated large-scale cluster state, is a promising candidate for practical, scalable, universal, and fault-tolerant quantum computation. In this work, based on our compact and scalable scheme of generating a two-dimensional spatiotemporal cluster state, a complete architecture including cluster state preparation, gate implementations, and error correction, is proposed. First, a scheme for generating two-dimensional large-scale continuous-variable cluster state by multiplexing both the temporal and spatial domains is proposed. Then, the corresponding gate implementations by gate teleportation are discussed and the actual gate noise from the generated cluster state is considered. After that, the quantum error correction can be further achieved by utilizing the square-lattice Gottesman-Kitaev-Preskill (GKP) code. Finally, a fault-tolerant quantum computation can be realized by introducing bias into the square-lattice GKP code (to protect against phase-flip errors) and concatenating a repetition code (to handle the residual bit-flip errors), with a squeezing threshold of 12.3 dB. Our work provides a possible option for a complete fault-tolerant quantum computation architecture in the future.
\end{abstract}
\begin{document}

\flushbottom
\maketitle

\thispagestyle{empty}

\section*{Introduction}

In measurement-based quantum computation (MBQC), gate implementations can be flexibly achieved by measurements with adaptively chosen basis. Continuous-variable (CV) MBQC, which requires deterministically generated large-scale cluster state, is a promising candidate for practical, scalable, universal, and fault-tolerant quantum computation\cite{Takeda(2019),Pfister(2019),Fukui(2021),furusawa(2024)PhysRevA.109.040101}. In general, a complete CV MBQC architecture mainly includes three steps: cluster state generation, universal quantum computation, and (fault-tolerant) quantum error correction.

As the prerequisite of CV MBQC, large-scale CV cluster state must be prepared first. In the past ten years, different kinds of large-scale CV cluster states have been experimentally generated by time\cite{yokoyama(2013),yoshikawa(2016),asavanant(2019),larsen(2019)} and frequency\cite{Pysher(2011),Chen(2014)} multiplexing. Also many theoretical schemes for generating large-scale cluster states by multiplexing time and frequency\cite{alexander(2016),wu(2020),Du(2022)}, time and space\cite{Barros(2020)controlled}, frequency and space\cite{zhangJing(2017),yangrg(2020)}, have been proposed recently. Considering the structure of the generated cluster states, at least two dimensions are required to achieve universal quantum computation in CV MBQC\cite{Van(2006)Universal}, one for computation and another for manipulation. 

After preparing the key quantum resource, large-scale cluster state(s), one can proceed to the gate implementation step. There are abundant theoretical demonstrations of single-mode Gaussian\cite{alexander2014noise,yokoyama(2013)}, two-mode Gaussian\cite{alexander(2016),alexander(2018),Alexander(2016)computationscheme,Alexander(2016)FlexibleQC,Larsen(2021)PRXQuantum} and single-mode non-Gaussian\cite{Furusawa(2021)non-Gaussian,Eaton(2019)Non-Gaussian} gates, by projection measurements based on gate teleportation or macronode structure. In experiment, single-mode Gaussian operation and further scalable quantum operations were demonstrated\cite{Asavanant(2020)One-hundred}, also a small quantum circuit consisting of 10 single-mode and 2 two-mode gates was executed on a three-mode input state\cite{Larsen(2020)quantum-computing}, both multiplexing in the time domain. 

Unfortunately, during gate implementations, Gaussian noise is inevitably added to the quantum information during computation, because only finite squeezing can be achieved in labs. To combat this noise, one can encode logical qubits into physical bosonic modes, using the Gottesman-Kitaev-Preskill (GKP) code\cite{Gottesman(2000)GKPcoding}. However, the GKP code can only correct the small displacement errors (usually less than $\sqrt{\pi}/2$) in phase space, and the large displacement errors are converted into Pauli errors in the encoded qubits\cite{MenicucciPRL(2014),Fukui(2017)REPETITIONCODE,KyungjooPRL(2020)}. Furthermore, residual errors can be further treated by combining GKP code with certain higher-level qubit error correction codes, such as the repetition code\cite{Fukui(2017)REPETITIONCODE,Stafford(2023)BiasedPhysRevA}, the [[4, 2, 2]] code\cite{Fukui(2018)422CODE}, or the surface code\cite{Fukui(2017)SURFACECODE,Vuillot(2019)SURFACE,Noh(2019)SURFACE}, towards fault-tolerant MBQC. 

A fault-tolerant MBQC scheme using the surface-GKP code was proposed and the corresponding squeezing threshold was 18.6 dB\cite{Noh(2019)SURFACE}. A more recent proposal introduced a fault-tolerant MBQC architecture based on a specially prepared three-dimensional (3D) cluster state using topological surface-4-GKP was proposed and the corresponding squeezing threshold was 12.7 dB\cite{Larsen(2021)PRXQuantum}. Another fault-tolerant MBQC architecture based on a Raussendorf-Harrington-Goyal cluster state was also proposed and the corresponding squeezing threshold was 13.6 dB\cite{TzitrinFaultTolerantPRXQuantum}. In this work, we propose a compact and scalable scheme of generating a large-scale bilayer-square-lattice-structured two-dimensional (2D) spatiotemporal cluster state. By combining biased-GKP code with repetition code, we validate a complete fault-tolerant CV MBQC architecture, considering the actual gate noise in the quantum gate implementations, the finite squeezing in the preparation of data and ancillary GKP states. 

\section{The complete fault-tolerant CV quantum computation architecture}
\begin{figure}[ht]
\centering
\includegraphics[height=0.25\textheight,width=0.7\linewidth]{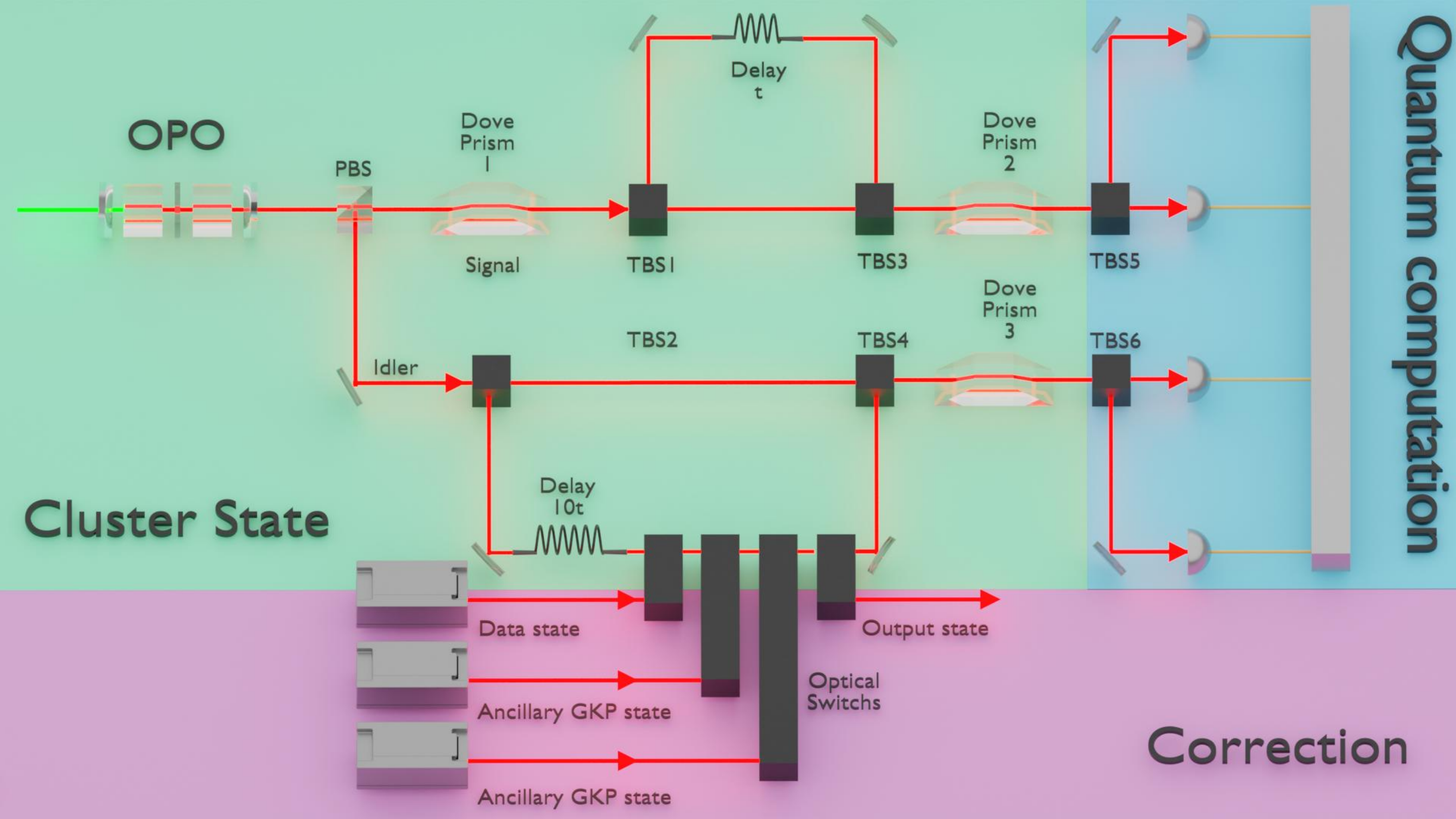}
\caption{The schematic diagram  of the complete continuous-variable quantum computation architecture.}
 \label{p1}
\end{figure}

To give an overall impression, a complete CV quantum computation architecture is shown in Fig.\ref{p1}, which consists of three parts: cluster state generation (green area), quantum computation (blue area), and quantum error correction (pink area). In the cluster state generation part, a cluster state with entanglement structure of bilayer square lattice is generated by multiplexing in both the temporal and spatial domains, which only requires one optical parametric oscillator (OPO). In the quantum computation part, single-mode and two-mode gates, which are necessary for universal quantum computation, can be efficiently implemented by homodyne detecting each generated entangled modes with proper measurement angle. Using optical switches\cite{larsen2019switch,Furusawa.switch(2019)}, the input data states and ancillary GKP states can be injected, and the result states can be output after a gate teleportation process. In the quantum correction part, the result states are projected onto the ancillary GKP states $\left|+\right\rangle_{\mathrm{GKP(R)}}$ and $\left|0\right\rangle_{\mathrm{GKP(R)}}$ through the $\hat{C}_{\mathrm{X}}(1)$ and $\hat{C}_{\mathrm{Z}}(1)$ gates, respectively. The ancillary GKP states are injected into the generated 2D cluster state during this process and the results of measuring the ancillary GKP states are considered in error correction.  In addition, fault-tolerant quantum computation can be further achieved by combining biased-GKP code with repetition code. All the details of each part will be given in the following sections.

\section{\label{sec:cluster state}Generation of the spatiotemporal cluster state }
\subsection{Scheme for generating the 2D cluster state}
\begin{figure}[ht]	\includegraphics[height=0.4\textheight,width=0.7\linewidth]{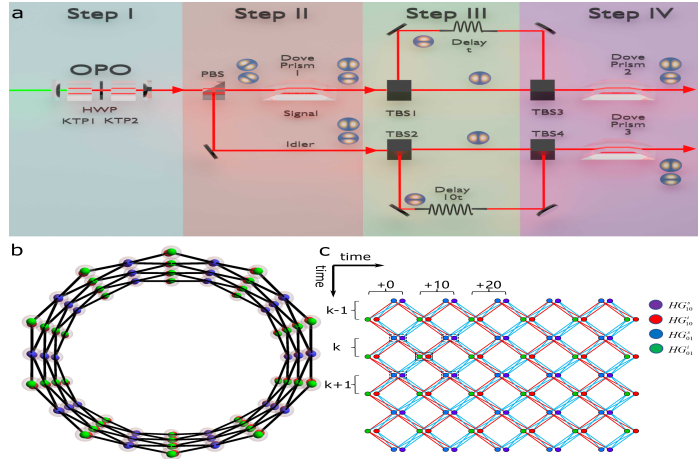}
\centering
	\caption{(a) The scheme for generating 2D spatiotemporal cluster state. (b) Whole structure.  (c) Detailed unrolled structure.}
 \label{p2}
\end{figure}

As is shown in Fig.\ref{p2}(a), the 2D cluster state can be generated by 4 steps. In step I, the pumped beam composed of Hermite Gaussian (HG) modes HG$_{02}$ and HG$_{20}$ is injected into the cavity, in which there is a pair of $\alpha$-cut type-II KTP crystals for simultaneous resonance of all modes and a half-wave plate (HWP) for Gouy-phase compensation. Then two pairs of entangled HG$_{01}$ and HG$_{10}$ modes with orthogonal polarization, i.e., HG$^{i}_{10}$, HG$^{s}_{10}$, HG$^{i}_{01}$, and HG$^{s}_{01}$ ($i$ and $s$ means idler and signal, respectively), are generated\cite{dos2009continuous,liu2014experimental}. The interaction Hamiltonian of the system is  
\begin{equation}
\hat{H}=i\hbar\xi\sum\mathbf{G} \left\{\hat{a}_{10}^{s\dagger }\hat{a}_{10}^{i\dagger}+\hat{a}_{01}^{s\dagger }\hat{a}_{01}^{i\dagger}+\mathrm{H.c.}\right\},
\end{equation}
where $\xi$ is the second-order nonlinear coefficient, and $\mathbf{G}$ is the adjacent matrix of $\mathrm{H}$ graph, whose matrix element is $1$ when the parametric process exists, and $0$ otherwise. $\hat{a}_{l}^{j\dagger}$ and $a_{l}^{j}$  ($j\in \left \{ i,s \right \}$ represents the polarization of the mode, $l\in \left \{01,10 \right \}$ represents the spatial mode) are the creation and annihilation operators of the fundamental fields, respectively. According to the properties of HG modes, one can get a square entangled state of (HG$^{s}_{45^{\circ}}$, HG$^{s}_{135^{\circ}}$, HG$^{i}_{10}$, HG$^{i}_{01}$), by coupling HG$^{s}_{10}$ and HG$^{s}_{01}$ modes as\cite{cai2018generation},
\begin{eqnarray}
\mathrm{HG}^{s}_{45^{\circ}}=\left ( \mathrm{HG}^{s}_{01}+\mathrm{HG}^{s}_{10}\right )/\sqrt{2},\quad \mathrm{HG}^{s}_{135^{\circ}}=\left (\mathrm{HG}^{s}_{01}-\mathrm{HG}^{s}_{10}\right )/\sqrt{2}. 
\label{HG}
\end{eqnarray} 
This transformation can be considered as a beam splitter transformation. The continuous-wave signal and idler beams can be divided into time bins of time period $t$, where $1/t$ is narrower than the bandwidth of the OPO cavity. Thus, a series of square entangled states separated by time interval $t$ can be deterministically created, and square entangled state in each time bin is independent of each other.

In step II, the square entangled state is divided into two parts: (HG$^{s}_{45^{\circ}}$, HG$^{s}_{135^{\circ}}$, horizontal polarization)  and (HG$^{i}_{01}$, HG$^{i}_{10}$, vertical polarization), by a polarized beam splitter (PBS).  Then the signal beams (HG$^{s}_{45^{\circ}}$, HG$^{s}_{135^{\circ}}$) are spatially rotated into (HG$^{s}_{10}$, HG$^{s}_{01}$) by a dove prism. Thus the entangled modes become (HG$^{s}_{10}$, HG$^{s}_{01}$, HG$^{i}_{10}$, HG$^{i}_{01}$). In step III, HG$^{l}_{10}$ and HG$^{l}_{01}$ modes in signal and idler beams are separated by transverse-mode beam splitters (TBSs), and then HG$^{s}_{01}$ and HG$^{i}_{01}$ are delayed by $t$ and $\mathrm{N}t$ ($\mathrm{N}$ can take arbitrary integer, here we set $\mathrm{N}=10$, as an example), respectively. After the fiber delay, the HG$^{s}_{01}$ mode of $k$ time is synchronized with the HG$^{s}_{10}$ mode of ($k+1$) time, while the HG$^{i}_{01}$ mode of $k$ time is synchronized with the HG$^{i}_{10}$ mode of ($k+10$) time. The delay lines multiplex these modes in time so that they form a 2D (but still disconnected) grid layout with cylindrical boundary.

In step IV, the staggered HG$^{l}_{01}$ and HG$^{l}_{10}$ modes are coupled by a spatial ``beam splitter'' which consists of a TBS and a dove prism. Thus, a series of square entangled states, which form a 2D grid layout with cylindrical boundary, are connected at this step. The generated 2D cluster state has the continuous cylindrical structure, as is shown in Fig.2(b).  Here each node (transparent gray ball) in the structure contains two modes: (HG$^{i}_{01}$ (green), HG$^{i}_{10}$ (red)), or (HG$^{s}_{01}$ (blue), HG$^{s}_{10}$ (purple)), and each edge (black line) represents a dual-rail structured connection. Fig.2(c) gives the detailed graph of unrolling this cylindrical structure, i.e., the bilayer square lattice structure, which is a universal resource for quantum computation. Note that the delay between temporal modes arriving at each detector is set by the shorter delay $t$. The longer delay line $\mathrm{N}t$ determines the circumference of the cylinder. Therefore, an arbitrary large 2D cluster state can be generated by choosing proper longer delay $\mathrm{N}t$, which is limited by the coherence time of the light source.

\subsection{Nullifier and VLF full-inseparability criterion}
To analyze the squeezing requirement of the 2D entangled state, we need to discuss the nullifier based on the van Loock–Furusawa (VLF) full-inseparability criterion, which can be understood as translation of the positive partial transpose condition into the form of the uncertainty relation\cite{Furusawavlf(2003)}. 

When the $\mathbf{G}$ matrix is self-inverse and bipartite, the complex adjacency matrix $\mathbf{Z}$ of Gaussian pure state can be written as\cite{Menicucci(2010)GraphicalCF}
\begin{equation}
    \mathbf{Z} =i\cosh\left ( 2r \right )\mathbf{I}-i\sinh \left ( 2r\right ) \mathbf{G},  
\end{equation}
where $\mathbf{I}$ represents an identity matrix and $r$ is the squeezing parameter. This $\mathrm{H}$ graph state can be transformed into a cluster state by operating a $\pi/{2}$ phase-space rotation on half of its modes or a $\pi/{4}$ phase-space rotation on all of its modes. Since this phase-space rotation can be absorbed into the measurement basis when measuring each mode of the state, a self-inverse bipartite $\mathrm{H}$ graph state can be considered as a cluster state. Therefore, the generated entangled state can be considered as a cluster state.

An $n$-mode Gaussian pure state $\left|\psi\right\rangle$ can be efficiently characterized by $n$ independent nullifiers, which are linear combinations of the quadrature operators, with $\left|\psi\right\rangle$ corresponding to their mutual zero-eigenstate. Nullifiers also play an important role in verifying genuine multipartite inseparability for cluster states transformed from self-inverse bipartite $\mathrm{H}$ graph states. Such cluster states are approximately nullified by linear combinations of quadratures that are either position- or momentum- type. For our 2D large-scale spatiotemporal cluster state, the nullifiers consist of six modes can be obtained as,
\begin{eqnarray}
    N_{k}^{x_{1}}=\left (\hat{x}_{10,k}^{i}-\hat{x}_{01,k}^{i}\right )-\left ( \hat{x}_{10,k+1}^{s}+\hat{x}_{01,k+1}^{s} \right )/\sqrt{2}
    +\left ( \hat{x}_{10,k}^{s}-\hat{x}_{01,k}^{s} \right )/\sqrt{2},
\end{eqnarray}
\begin{eqnarray}
N_{k}^{x_{2}}=\left(\hat{x}_{10,k+10}^{i}+\hat{x}_{01,k+10}^{i}\right )-\left ( \hat{x}_{10,k+1}^{s}+\hat{x}_{01,k+1}^{s} \right )/\sqrt{2}
    -\left(\hat{x}_{10,k}^{s}-\hat{x}_{01,k}^{s}\right )/\sqrt{2},
\end{eqnarray}
\begin{eqnarray}
     N_{k}^{p_{1}}=\left (\hat{p}_{10,k}^{i}-\hat{p}_{01,k}^{i}\right )+\left ( \hat{p}_{10,k+1}^{s}+\hat{p}_{01,k+1}^{s} \right )/\sqrt{2}
    -\left ( \hat{p}_{10,k}^{s}-\hat{p}_{01,k}^{s} \right )/\sqrt{2},
\end{eqnarray}
\begin{eqnarray}
     N_{k}^{p_{2}}=\left (\hat{p}_{10,k+10}^{i}+\hat{p}_{01,k+10}^{i}\right )+\left ( \hat{p}_{10,k+1}^{s}+\hat{p}_{01,k+1}^{s} \right )/\sqrt{2}
    +\left ( \hat{p}_{10,k}^{s}-\hat{p}_{01,k}^{s} \right )/\sqrt{2}.
\end{eqnarray}
In fact, nullifiers span a vector space, and any linear combination of nullifiers is still a nullifier. Therefore the generated 2D cluster state can also be specified with the following set of nullifiers,
\begin{eqnarray}
(N_{k+10}^{x_{1}}+N_{k}^{x_{2}})/{2}=\hat{x}_{10,k+10}^{i}-\left ( \hat{x}_{10,k+11}^{s}+\hat{x}_{01,k+11}^{s} \right )/2\sqrt{2} -\left ( \hat{x}_{01,k+10}^{s}-\hat{x}_{10,k+10}^{s} \right )/2\sqrt{2}\nonumber\\
    -\left ( \hat{x}_{10,k+1}^{s}+\hat{x}_{01,k+1}^{s} \right )/2\sqrt{2}
    -\left ( \hat{x}_{10,k}^{s}-\hat{x}_{01,k}^{s} \right )/2\sqrt{2},
\end{eqnarray}
\begin{eqnarray}
(-N_{k+10}^{x_{1}}+N_{k}^{x_{2}})/{2}=\hat{x}_{10,k+10}^{i}+\left ( \hat{x}_{10,k+11}^{s}+\hat{x}_{01,k+11}^{s} \right )/2\sqrt{2}
    -\left ( \hat{x}_{10,k+10}^{s}-\hat{x}_{01,k+10}^{s} \right )/2\sqrt{2}\nonumber\\-\left ( \hat{x}_{10,k+1}^{s}+\hat{x}_{01,k+1}^{s} \right )/2\sqrt{2}
    -\left ( \hat{x}_{10,k}^{s}-\hat{x}_{01,k}^{s} \right )/2\sqrt{2},
\end{eqnarray}
\begin{eqnarray}   (N_{k+10}^{p_{1}}+N_{k}^{p_{2}})/{2}=\hat{p}_{10,k+10}^{i}+\left ( \hat{p}_{10,k+11}^{s}+\hat{p}_{01,k+11}^{s} \right )/2\sqrt{2}
    +\left ( \hat{p}_{01,k+10}^{s}-\hat{p}_{10,k+10}^{s} \right )/2\sqrt{2}\nonumber\\+\left ( \hat{p}_{10,k+1}^{s}+\hat{p}_{01,k+1}^{s} \right )/2\sqrt{2}
    +\left ( \hat{p}_{10,k}^{s}-\hat{p}_{01,k}^{s} \right )/2\sqrt{2},
\end{eqnarray}
\begin{eqnarray}    (-N_{k+10}^{p_{1}}+N_{k}^{p_{2}})/{2}=\hat{p}_{10,k+10}^{i}-\left ( \hat{p}_{10,k+11}^{s}+\hat{p}_{01,k+11}^{s} \right )/2\sqrt{2}
    +\left ( \hat{p}_{10,k+10}^{s}-\hat{p}_{01,k+10}^{s} \right )/2\sqrt{2}\nonumber\\+\left ( \hat{p}_{10,k+1}^{s}+\hat{p}_{01,k+1}^{s} \right )/2\sqrt{2}
    +\left ( \hat{p}_{10,k}^{s}-\hat{p}_{01,k}^{s}\right)/2\sqrt{2},
\end{eqnarray}
these nullifiers match well with the entanglement structure in Fig.\ref{p2}(c). When the red (HG$^{i}_{10}$) and green (HG$^{i}_{01}$) balls in the solid rectangular box are focused on, it is found that they connect with other four sets of blue (HG$^{s}_{01}$) and purple (HG$^{s}_{10}$) balls in the dashed rectangular boxes. The nullifiers of quadrature amplitude and phase of the red and green balls are corresponding to the Eq.(8)-(11), respectively.
\begin{figure}[ht]
    \centering
    \subfigure[]{\label{1}\includegraphics[width=0.35\textwidth]{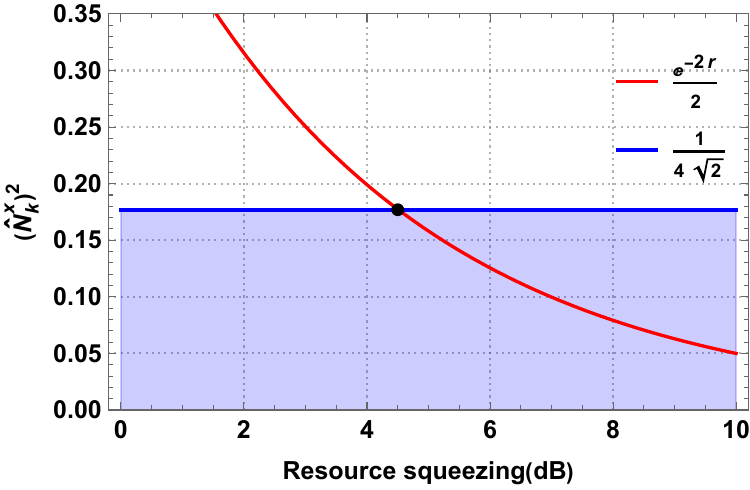}}\subfigure[]{\label{2}\includegraphics[width=0.35\textwidth]{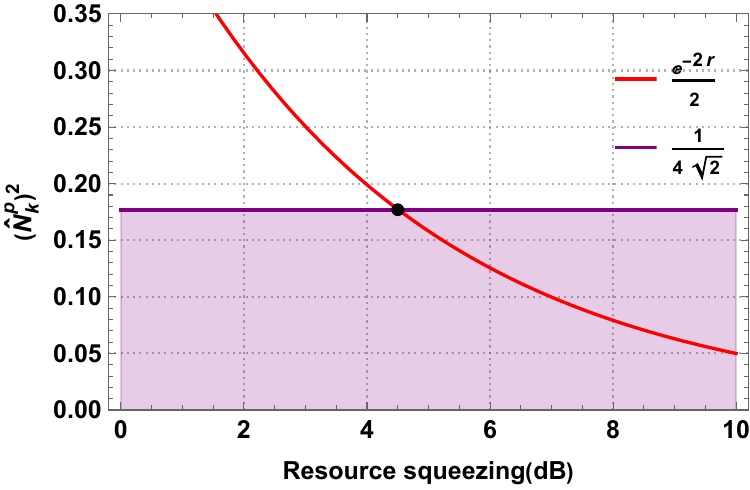}}
    \caption{The nullifier of quadrature $\hat{x}$ (a) and $\hat{p}$ (b) versus the squeezing parameter, for the generated 2D cluster state. }
    \label{p3}
\end{figure}

For VLF criterion, a minimum unit of the generated 2D cluster state are divided into two sets, $s_{1}$ and $s_{2}$, and a linear combination of quadratures can be defined\cite{yokoyama(2013)},
\begin{equation}
    \hat{u}=\sum_{j\in s_{1}\cup s_{2}}h_{j}\hat{x}_{j},\quad \hat{v}=\sum_{j\in s_{1}\cup s_{2}}g_{j}\hat{p}_{j},
\end{equation}
choose the required the linear combination of quadratures from Eq.(4)-(8). The VLF criterion can be expressed as
\begin{equation}
    \left\langle\Delta\hat{u}^{2}\right\rangle + \left\langle\Delta\hat{v}^{2}\right\rangle< \left|\sum_{j\in s_{1}}h_{j}g_{j}\right|+\left|\sum_{j\in s_{2}}h_{j}g_{j}\right|,
\end{equation}
when the inequality is satisfied, two sets are inseparable. For a minimum unit of the 2D cluster state, there are 6 basic modes and 31 possible bipartitions, based on which the following inequalities as a sufficient condition due to full-inseparability can be obtained as
\begin{equation}
\left \langle \left (\hat{N}_{k}^{x_{1}}\right )^2\right \rangle<1/{4\sqrt{2} },\qquad  \left \langle \left (\hat{N}_{k}^{p_{1}} \right )^2\right \rangle<1/{4\sqrt{2}}.
\end{equation}
While the variances of nullifiers $\hat{N}_{k}^{x_{1}}$ and $\hat{N}_{k}^{p_{1}}$ can be derived:
\begin{equation}
\begin{aligned}
\left \langle \left (\hat{N}_{k}^{x_{1}}\right )^2\right \rangle=e^{-2r}/2,\qquad
\left \langle \left (\hat{N}_{k}^{p_{1}} \right )^2\right \rangle=e^{-2r}/2.
\end{aligned} 
\end{equation}
Therefore, the above conditions can determine the required minimum squeezing for generating 2D cluster states. As is shown in Fig.(3), when the squeezing is larger than 4.5 dB, the sufficient condition for full-inseparability will be satisfied.

\section{Quantum computation based on 2D spatiotemporal cluster state}
\begin{figure}[ht]
\centering
\includegraphics[width=0.5\textwidth]{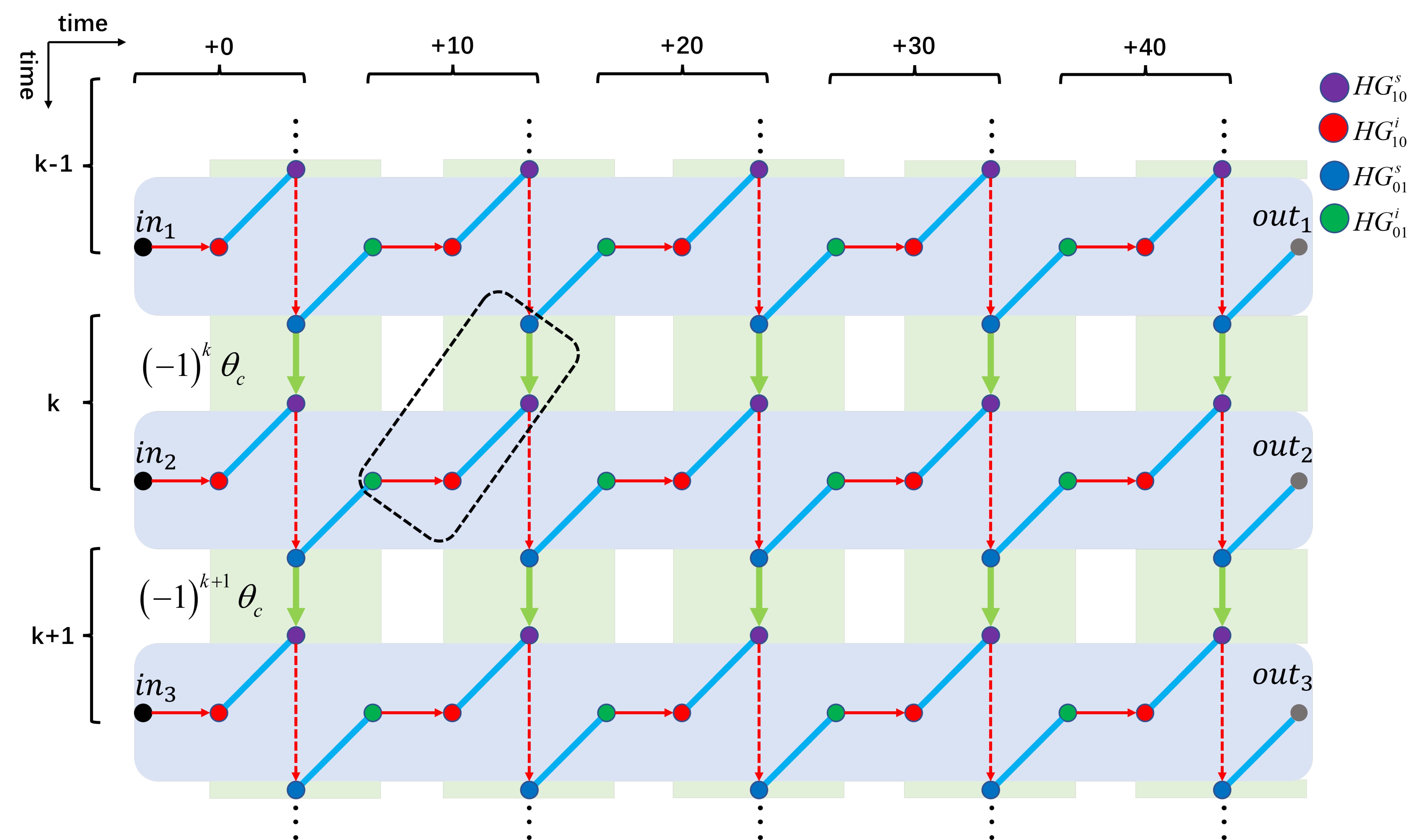}
\centering
\caption{Simplified graphical-calculus representation of quantum computation based on the bilayer square lattice CV cluster state. Here the blue lines correspond to the entanglement weights of $\tanh(2r)$.}
\label{p4}
\end{figure}

After generating a large-scale 2D cluster state, how to use it in MBQC is now under our consideration. To analyze the quantum computation scheme based on our generated 2D cluster state, the graphical-calculus representation with simplified entanglement structure is shown in Fig.\ref{p4}\cite{Menicucci(2010)GraphicalCF}. The resource squeezing is $\delta=e^{-2r}$, and the corresponding squeezing contained in the generated 2D cluster state is $\varepsilon=\operatorname{sech(2r)}$. Here the blue lines correspond to the entanglement weights of $\tanh(2r)$, the red dashed arrows represent the process of Eq.(2), while the red and green solid arrows represent the action of the spatial ``beam splitter'' for idler and signal beam, respectively. The four modes in the black dashed rectangle arrive the detectors simultaneously at $k+10$ time.

The input states shown by the black nodes can be switched into the circuit and teleported along the quantum wires (light-blue area) by using gate teleportation in sequence. The neighbor quantum wires can be connected through the control wires (light-green area). The modes in quantum wires that carry quantum information to be processed are called wire modes. And the modes in control wires that control the connectivity between the neighbor wires are called control modes. As is shown in Fig.(\ref{p4}), the input states can be teleported along Z-like quantum wires (along four spatial modes in different time intervals ) by using gate teleportation, which is different from the wires along dual-rail cluster states in Ref.\cite{alexander(2016)}.

\begin{figure}[ht]
\centering
\includegraphics[width=\textwidth]{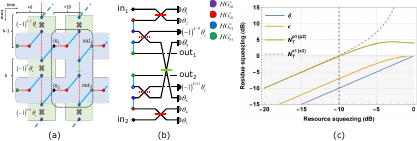}
\caption{The logic level (a) and the equivalent quantum circuit (b) of implementing two-mode gate quantum computation on the bilayer-square-lattice cluster state. (b) corresponds to the black dashed box part in (a). The green solid, red solid, and red dashed beam splitters correspond to the green solid, red solid, and red dashed arrows, respectively. (c) resource remaining squeezing after quantum gates operation.}
\label{P5}
\end{figure}

For single-mode gate-implementations, neighbor quantum wires must be decoupled from each other, i.e., the measurement angles of control modes HG$^{s}_{01,k}$ and HG$^{s}_{10,k}$ are set as $\left(-1\right)^{k}\theta_{c}$=$\pm \pi/{4}$. For two-mode gate-implementations, such as controlled-Z and controlled-X gates,  the measurement restrictions on certain control modes must be lifted\cite{alexander(2016),Menicucci(2010)GraphicalCF}, as is shown by the black dashed box part in Fig.\ref{P5}(a). The corresponding equivalent quantum circuit of the two-mode gate implementation is shown in Fig.\ref{P5}(b). Here the obtained controlled-Z (controlled-X) gate can be expressed as
\begin{eqnarray}
\left [\hat{R}\left(\mp 3\pi/{4}\right)\otimes\hat{R}\left(\pm \pi/{4}\right)\right]\hat{C}_{\mathrm{Z}}\left(g\right)=\left[\hat{R}\left(\pm3\pi/{4}\right ) \otimes\hat{R}\left(\mp\pi/{4}\right)  \right]\hat{C}_{\mathrm{X}}\left(g\right),   
\end{eqnarray}
with the chosen measurement angles
\begin{eqnarray}
\left ( \theta_{1},\theta_{2},\theta_{3},\theta_{4},\theta_{5},\theta_{6}\right )=
\left (-\frac{\pi }{8},\frac{3\pi }{8},-\frac{\pi }{8},\frac{3\pi }{8},\frac{\pi }{4}\pm arctan \frac{2}{g},\frac{\pi }{4}\mp arctan \frac{2}{g}\right ),
\label{CZGATEANGLE}
\end{eqnarray}
where $g$ is the tunable coupling strength of the controlled-Z (controlled-X) gate, which is followed by the byproduct (known, fixed) $\hat{R}\left (\mp3\pi/{4}\right ) \otimes\hat{R}\left(\pm\pi/{4}\right)$. By applying the inverse single-mode gates immediately after this gate, this byproduct part can be compensated in the next step.

However, Gaussian noise is inevitably added to the quantum information during computation due to infinite squeezing. When $g=1$, the resulting gate noise $\mathcal{N}$ for $\hat{C}_{\mathrm{Z}}(1)$ ($\hat{C}_{\mathrm{X}}(1)$) gate is
\begin{equation}
\mathcal{N} = \begin{pmatrix}
 \frac{\sqrt{2}}{-\Gamma^{2}} & \frac{\sqrt{2}}{2\Gamma^{2}} & -\frac{3\sqrt{2}}{4\Gamma} & -\frac{\sqrt{2}}{4\Gamma} & -\frac{1}{\Gamma} & -\frac{1}{2\Gamma} & 0 & 0 \\
 -\frac{\sqrt{2}}{2\Gamma^{2}} & \frac{\sqrt{2}}{\Gamma^{2}} & \frac{3\sqrt{2}}{4\Gamma} & -\frac{\sqrt{2}}{4\Gamma} & -\frac{1}{2\Gamma} & -\frac{1}{\Gamma} & 0 & 0 \\
 0 & -\frac{\sqrt{2}}{2} & \frac{\sqrt{2}\Gamma}{4} & \frac{\sqrt{2}\Gamma}{4} & -\Gamma & \frac{\Gamma}{2} & \sqrt{2} & 0 \\
 \frac{\sqrt{2}}{2} & 0 & -\frac{\sqrt{2}\Gamma}{4} & \frac{\sqrt{2}\Gamma}{4} & \frac{\Gamma}{2} & -\Gamma & 0 & \sqrt{2}
\end{pmatrix},
\end{equation}
where $\Gamma=\tanh(2r)/\sqrt{2}$, the gate noises that added to the output quadratures $\hat{x}_{1}$($\hat{x}_{2}$) and $\hat{p}_{1}$($\hat{p}_{2}$) are $\mathcal{N}_{x_{1}}\varepsilon$ ($\mathcal{N}_{x_{2}}\varepsilon$), and $\mathcal{N}_{p_{1}}\varepsilon$ ($\mathcal{N}_{p_{2}}\varepsilon$), respectively, with
\begin{eqnarray}
\mathcal{N}_{{x}_{1}} =\mathcal{N}_{{x}_{2}}=\frac{5}{2} \left (\frac{1}{tanh^{4}(2r)}+\frac{1}{tanh^{2}(2r)}\right ),\quad
	\mathcal{N}_{{p}_{1}}= \mathcal{N}_{{p}_{2}} =\frac{5}{2} \left ( tanh^{2}(2r)+1 \right ). 
\end{eqnarray}

To quantify the effect of the quantum gate noises, the residual squeezing after implementing the two-mode gate, compared with the resource squeezing and the squeezing contained in the generated 2D cluster state, are shown in Fig.\ref{P5}(c). Note that the squeezing in the cluster state modes ($\varepsilon=\operatorname{sech(2r)}$) is reduced about 3 dB compared with the resource squeezing ($\delta=e^{-2r}$) in high squeezing level, which is the cost of preparing the cluster state. For example, when the resource squeezing is 10 dB, the corresponding squeezing contained in the generated 2D entangled state is 7 dB, the residual squeezing of the two-mode gates is 0 dB. This means that a resource squeezing of 10 dB can only support one two-mode gate at most, which is far from enough for quantum computation.

\section{Quantum error correction}
It is now clear that quantum computation based cluster state will inevitably suffer from gate noise that accumulates throughout the computation. The effects of gate noise are modeled by the application of an additive Gaussian bosonic channel on the cluster state\cite{Hallrandomnoise(1994)}:
\begin{equation} N\left[\sigma\right]\left(\hat{\rho}\right)=\int\frac{d^{2}\alpha}{\pi\sigma^{2}}exp\left[-\frac{\left|\alpha\right|}{\sigma^{2}}\right]\hat{D}\left(\alpha\right)\hat{\rho}\hat{D}^{\dagger}\left(\alpha\right), 
\end{equation}
Here, $\hat{D}\left(\alpha\right)$ is the displacement operator and $\sigma$ is the standard deviation of the random displacement. More explicitly, the action of $N\left[\sigma\right]$ transforms the quadrature amplitude and phase as follows in the Heisenberg picture:
\begin{eqnarray}
\hat{x}\rightarrow\hat{x}+\xi_{x},\quad
\hat{p}\rightarrow\hat{p}+\xi_{p},
\end{eqnarray}
the gate noises add shift errors $\xi_{x}$ and $\xi_{p}$ to the quadrature amplitude and phase, respectively. Therefore, the efficient GKP encoding is chosen to correct quadrature error between implemented gates. In fact, GKP state is often considered not only as a non-Gaussian resource but also as an ancillary resource for error correction\cite{Gottesman(2000)GKPcoding}. 
\begin{figure}[ht]
\includegraphics[width=\textwidth]{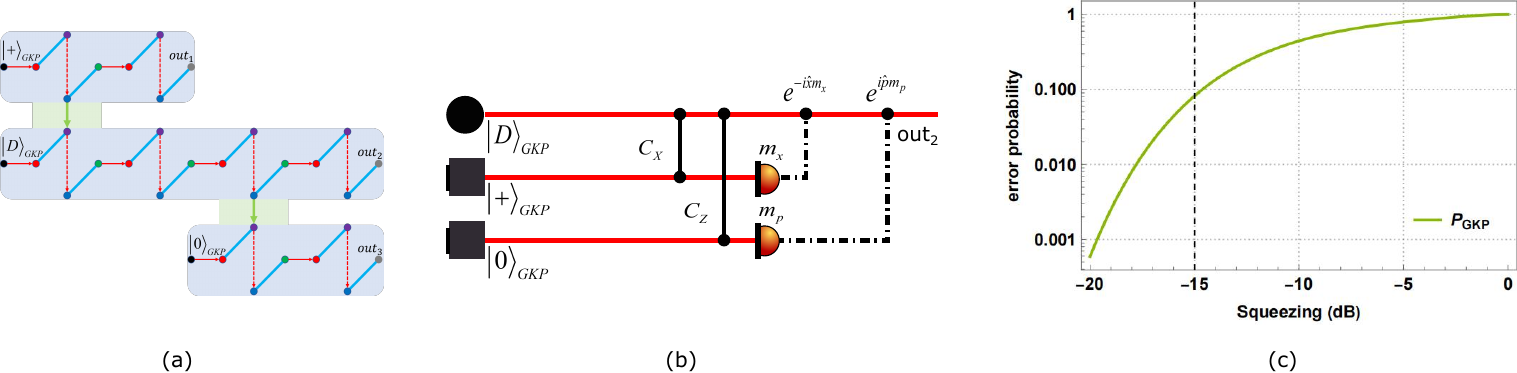}
\centering
	\caption{The logic level (a) and the equivalent quantum circuit (b) of GKP quantum correction for correcting gate noise after each implemented gate. (c) error probability for quantum error correction using GKP code.}
 \label{p6}
\end{figure}

\subsection{GKP code}
Here, we consider the GKP state which can be defined as the two-dimensional subspace of a bosonic Hilbert space that is stabilized by the two stabilizers
\begin{equation}
	\hat{S}_{x}\equiv\exp\left [ i2\sqrt{\pi/R}\hat{q}\right],\qquad \hat{S}_{p}\equiv\exp\left [ i2\sqrt{\pi R}\hat{p}\right].  
\end{equation}
The computational basis states of the ideal GKP states are given by
\begin{eqnarray}
\left |0\right \rangle_{\mathrm{GKP}\left(\mathrm{R}\right)}=\sum_{n\in\mathbb{Z}}\left |\hat{x}=2n\sqrt{\pi/R }\right \rangle,\qquad
	\left |1\right \rangle_{\mathrm{GKP}\left(\mathrm{R}\right)}=\sum_{n\in\mathbb{Z}}\left |\hat{x}=\left (2n+1\right )\sqrt{\pi/R }\right \rangle,  
\end{eqnarray}
and the alternative basis states are
\begin{eqnarray}
\left |+\right \rangle_{\mathrm{GKP}\left(\mathrm{R}\right)}=\sum_{n\in\mathbb{Z}}\left|\hat{p}=2n\sqrt{\pi R}\right \rangle, \qquad
	\left |-\right \rangle_{\mathrm{GKP}\left(\mathrm{R}\right)}=\sum_{n\in\mathbb{Z}}\left|\hat{p}=\left (2n+1\right )\sqrt{\pi R}\right \rangle, 
\end{eqnarray}
where $\mathrm{R}\in\mathbb{R}>0$ is the aspect ratio of the lattice, with $\mathrm{R}=1$ corresponding to the square-lattice GKP states. Ideal GKP states, including the computational basis states, are unphysical states. In our architecture, we consider the physical approximate GKP states, which replacing each $\delta$ spike in the quadrature wave function by a symmetric Gaussian spike with a variance of $\sigma$, and the whole states are modulated by a Gaussian envelope\cite{campagne2020GKP,fluhmann2019GKP,konno2024GKP}.  

How the ancillary GKP states are used to correct quadrature error of output data GKP state without destroying information in our architecture, is shown in Fig.\ref{p6}(a). The corresponding equivalent quantum circuit is shown in Fig.\ref{p6}(b). The success probability of the GKP error correction can be obtained as
\begin{eqnarray}
p_{\mathrm{succ}}^{R}=\sum_{n_{x},n_{p}\in2\mathbb{Z}}\int_{\left(n_{x}-\frac{1}{2}\right)\sqrt{\frac{\pi}{R}}}^{\left(n_{x}+\frac{1}{2}\right)\sqrt{\frac{\pi}{R}}}d\xi_{x}f_{\sigma_{x}}(\xi_{x})
\int_{\left(n_{p}-\frac{1}{2}\right)\sqrt{\pi R}}^{\left(n_{p}+\frac{1}{2}\right)\sqrt{\pi R}}d\xi_{p}f_{\sigma_{p}}(\xi_{p}),
\label{P-GKP}
\end{eqnarray}
where $f_{\sigma}(\xi)=\frac{1}{\sqrt{2\pi\sigma^{2}}}e^{\left [-\frac{\xi^{2}}{2\sigma^{2}}\right]}$ is Gaussian distribution
function with variance $\sigma^{2}$. Note that the quadrature variance of Gaussian spikes in the approximate GKP state after GKP error correction is the main factor that affecting the error probability. In other works, they often consider only the finite squeezing in the preparation of data and ancillary GKP states, the gate noise variance $\sigma^{2}_{n}$ is not considered\cite{Bourassa2021blueprintscalable,TzitrinFaultTolerantPRXQuantum} or is taken to be $\sigma^{2}_{n}=e^{-2r}$\cite{Larsen(2021)PRXQuantum}. Here, we consider $\sigma_{x(p)}^{2}=\sigma^{2}_{0}+\mathcal{N}_{x(p)}\varepsilon+\sigma^{2}_{\mathrm{A}}$, which includes not only the actual gate noise $\sigma^{2}_{n}=\mathcal{N}\varepsilon$ in the above gate implementations, but also the finite squeezing in the preparation of data ($\sigma^{2}_{0}$) and ancillary ($\sigma^{2}_{\mathrm{A}}$) GKP states. In our noise model, measure noise and idle noise are ignored, since measurements have been implemented with a pretty high efficiency and all modes are measured in the optical MBQC platform.

Here, the quadrature squeezing of the data state $\sigma^{2}_{0}$, the ancillary GKP state $\sigma^{2}_{\mathrm{A}}$ and the resource squeezing $\delta=e^{-2r}$ are assumed to be equal. When $\mathrm{R}=1$, the error probabilitie versus squeezing are shown in Fig.\ref{p6}(c). It is clear that the higher the squeezing level the lower the error probability. For example, if squeezing is 15 dB, which is the present maximum squeezing obtained in experiment\cite{Vahlbruch(2016)}, the error probabilities is about 0.08. Although GKP encoding is effective to most Gaussian random shift errors, it is still far from enough for practical quantum computation. Therefore, to further reduce the error probability, the Pauli errors added to the data state by GKP encoding must be corrected, for achieving fault-tolerant quantum computation.

\section{Fault-tolerant quantum computation}
To reduce the error probability to a necessary ultra-low level, the square-lattice GKP state is extended to rectangular-lattice GKP state, i.e., a bias is introduced, to protect against phase-flip errors. Combining with a $n$-qubit repetition encoding, bit-flip errors can be corrected as well. The biased GKP repetition is proposed by Stafford\cite{Stafford(2023)BiasedPhysRevA}. Here we generalize it in our CV MBQC architecture. It is worth mentioning that other error correction codes, such as GKP-surface codes\cite{Noh(2019)SURFACE} or GKP-color codes\cite{JiaxuancolorGKP}, can also be chosen instead. 

\subsection{Biased GKP code}
Up to now, square-lattice GKP states are most widely studied due to their balanced protection in each quadrature. However, in principle, other alternative lattice grid states can also be defined\cite{Stafford(2023)BiasedPhysRevA}. Here rectangular-lattice GKP state is taken into account, and the balanced protection is broken, i.e., biasing the Pauli errors of the data state. The probabilities of Pauli X and Pauli Z errors are  
\begin{eqnarray}
    p_{\mathrm{X}} = 
\left( 1 - \sum_{n_{x} \in 2\mathbb{Z}} \int_{\left(n_{x} - \frac{1}{2}\right) \sqrt{\frac{\pi}{R}}}^{\left(n_{x} + \frac{1}{2}\right) \sqrt{\frac{\pi}{R}}} 
d\xi_{x} \, f_{\sigma_{x}}(\xi_{x}) \right) 
\sum_{n_{p} \in 2\mathbb{Z}} \int_{\left(n_{p} - \frac{1}{2}\right) \sqrt{\pi R}}^{\left(n_{p} + \frac{1}{2}\right) \sqrt{\pi R}} 
d\xi_{p} \, f_{\sigma_{p}}(\xi_{p}),  \\ 
p_{\mathrm{Z}} = \left( 1 - \sum_{n_{p} \in 2\mathbb{Z}} \int_{\left(n_{p} - \frac{1}{2}\right)\sqrt{\pi R}}^{\left(n_{p} + \frac{1}{2}\right)\sqrt{\pi R}} 
d\xi_{p} \, f_{\sigma_{p}}(\xi_{p}) \right) 
\sum_{n_{x} \in 2\mathbb{Z}} \int_{\left(n_{x} - \frac{1}{2}\right)\sqrt{\frac{\pi}{R}}}^{\left(n_{x} + \frac{1}{2}\right)\sqrt{\frac{\pi}{R}}} 
d\xi_{x} \, f_{\sigma_{x}}(\xi_{x}).
\end{eqnarray}
It is noted that biased GKP state has different tolerances for Pauli Z and Pauli X errors. While Pauli Z and Pauli Y errors can be suppressed by increasing $\mathrm{R}$, at the cost of increasing the Pauli X error rate, which can be solved by combining with a $n$-qubit bit-flip repetition encoding. 

\subsection{Concatenated repetition code}
In our scheme, a bit-flip repetition code is concatenated after biased GKP code. An $n$-qubit repetition code is given by 
\begin{equation}
    \left|\bar{0}\right\rangle=\left |0\right \rangle_{\mathrm{GKP}}^{\otimes n},\qquad \left|\bar{1}\right\rangle=\left |1\right \rangle_{\mathrm{GKP}}^{\otimes n}.
\end{equation}
For this repetition code, the probability of $j$ bit-flip errors obeys the binomial distribution
\begin{equation}
P\left ( j \right )= \left ( \begin{matrix}
 n\\
 j
\end{matrix} \right )\left (p_{\mathrm{X}}\right )^{j}\left ( 1-p_{\mathrm{X}} \right )^{n-j}.   
\end{equation}
The decoding process of bit-flip repetition code is a simple majority vote and it can correct up to $k$ bit-flip errors, where $k=\left\lfloor\left(n-1\right)/2\right\rfloor$. Therefore, the success probability $p\left(j\le k\right )$ of correcting bit-flips can be obtained as,
\begin{equation}
    P^{X}_{\mathrm{succ}}\left(j\le k\right)=\sum_{j=0}^{k}\binom{n}{j}\left (p_{\mathrm{X}}\right )^{j}\left(1-p_{\mathrm{X}}\right)^{n-j},
\end{equation}
which is just a partial binomial sum containing all combinations of outcomes.

The bit-flip repetition code does not have built-in protection against any phase-flip errors, and we cannot completely bias away quadrature phase displacement errors. Since even number of phase-flip errors will not change the logical state, here only odd number condition is concerned with. For an $n$-qubit code, the probability that even numbers of Pauli Z errors occurs, can be expressed by
\begin{eqnarray}
P^{Z}_{\mathrm{succ}}\left(2j\right)=\sum_{j=1}^{\left \lfloor n/2\right \rfloor }\binom{n}{2j}\left (p_{\mathrm{Z}}\right )^{2j}\left(1-p_{\mathrm{Z}}\right)^{n-2j}
=\left(1+\left(1-2p_{\mathrm{Z}}\right)^{n}\right)/{2}.
\end{eqnarray}
Note that the ancillary GKP states have the same aspect ratio as the data GKP states, and the weights of all two-mode controlled-Z and controlled-X gates are fixed at $g=1$.

\subsection{Calculation results}
We consider each error to give the overall code performance of the biaesd GKP repetition code. In our fault-tolerant quantum computation scheme, the inner rectangular-lattice GKP code is used to handle with the Gaussian random shift noise that acting on data GKP state by transforming Gaussian noise into Pauli errors, meanwhile the Pauli Z error is effectively suppressed. Then outer bit-flip repetition code is ready to combat the unsuppressed Pauli X error. Considering the residual Pauli Z errors and the uncorrected Pauli X errors, the overall code error probability is then given by\cite{Stafford(2023)BiasedPhysRevA}
\begin{equation}
    P_{e}=1-P^{X}_{\mathrm{succ}}\left(j\le k\right) P^{Z}_{\mathrm{succ}}\left(2j\right).
\end{equation} 
Note that $\mathrm{R}$ must be optimised depending on $n$.
\begin{figure}[ht]
	\includegraphics[width=\textwidth]{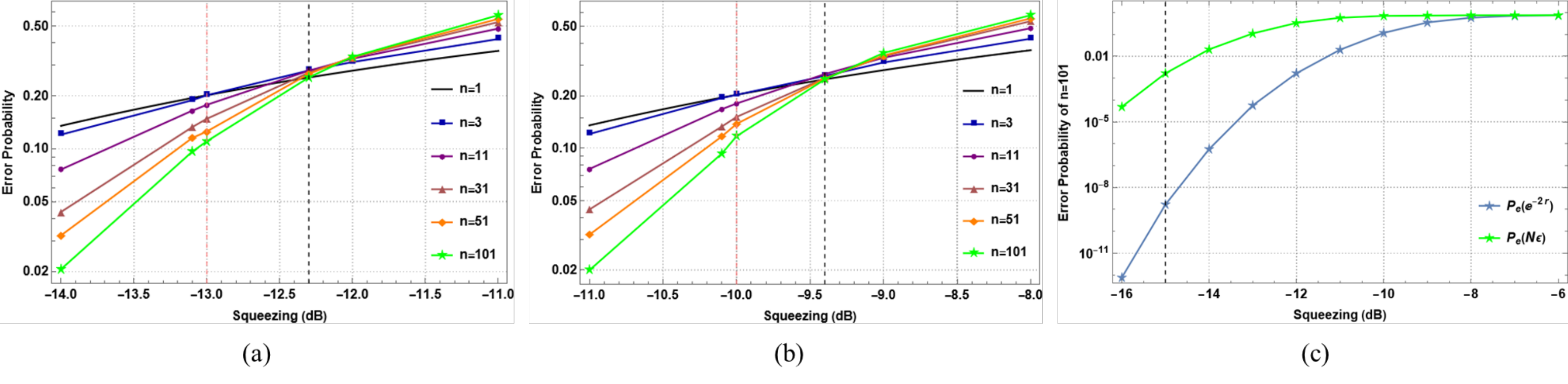}
 \centering
    \caption{Error probabilities ($P_{e}$) using the biased GKP repetition code, for $\sigma^{2}_{n}=\mathcal{N}\varepsilon$ (a), $\sigma_{n}=e^{-2r}$ (b), and compared results for n=101 (c).}
    \label{p7}
\end{figure}

In our scheme, the fault-tolerant threshold is 12.3 dB, as is shown in Fig.\ref{p7}(a). The figure can be categorised into three distinct regions that divided by the red and black dashed lines, according to the code behaviour. The intersection of $n$=1 and $n$=101 corresponds the fault-tolerant threshold. In the right region, the biased GKP repetition code performs worse than the no-bias (square) no-repetition condition ($n$=1).  In the middle region, the biased GKP repetition code starts to show the advantages and all repetition codes gradually perform better than the no-bias no-repetition condition. In the left region, the biased GKP repetition code shows absolute superiority, even for very small repetition numbers ($n$=3). It is found that the advantage of the biased GKP repetition code will be obvious if the squeezing level is high enough. Moreover, the error probability can be further deduced by increasing either the repetition number or the squeezing level. As is shown in Fig.\ref{p7}(b), the resulting squeezing threshold will be 9.4 dB, if considering $\sigma_{n}=e^{-2r}$ as in other MBQC schemes\cite{Larsen(2021)PRXQuantum,TzitrinFaultTolerantPRXQuantum}. In addition, the behaviors of the biased GKP repetition code ($n$=101) for $\sigma^{2}_{n}=\mathcal{N}\varepsilon$ and $\sigma_{n}=e^{-2r}$ are compared in Fig.\ref{p7}(c), which have at least two orders of magnitude improvement compared with that in Fig.\ref{p6}(c).

\section{Conclusion}
In this work, we investigate a complete CV quantum computation architecture, which includes 2D spatiotemporal cluster state generation, gate implementations, error correction, and fault tolerance (with squeezing threshold 12.3 dB). Here we obtain the squeezing threshold by taking into account the actual gate noise caused by finite squeezing of the generated 2D cluster state and GKP state in experiment. The computational performance can be further improved by increasing either the squeezing level or the repetition number, so one can finally obtain an ultra-low error probability by using the biased GKP repetition code, which has advantages of easier decoding and less stabilizer measurements requirement at the qubit level. In our next work, we would like to further optimize each steps in our quantum computation architecture, such as extending the cluster state from 2D to 3D, reducing the gate noise by streamlining the universal quantum gates\cite{BaragiolaStreamlined(2021)}, considering the analog information during GKP error correction\cite{Fukui(2017)REPETITIONCODE}, etc. Our work provides a potential pathway for realizing a fault-tolerant measurement-based CV quantum computation in experiment.

\section*{Data availability}
The datasets used and/or analysed during the current study available from the corresponding author on reasonable request.

\bibliography{sample}
\section*{Funding}
This work is supported by the National Key Research and Development Program of China (Grants No. 2021YFC2201802,  2021YFA1402002); the National Natural Science Foundation of China (Grants No. 11974225, No. 11874248 and No. 11874249), the National Key Laboratory of Radar Signal Processing (Grant No. JKW202401), and the Postgraduate Education Innovation Program of Shanxi Province (Grants No. 2024KY018).

\section*{Competing interests}
The authors declare no competing interests.
\end{document}